\documentclass[aps,twocolumn,secnumarabic,nobalancelastpage,nofootinbibt]{revtex4}

\usepackage{graphics}      
\usepackage{graphicx}      
\usepackage{longtable}     
\usepackage{url}           
\usepackage{bm,color}            
\usepackage{amssymb, amsmath, enumerate, theorem, epsfig,subfigure,setspace}
\usepackage{amsfonts}

\begin{document}

\title{Blockchain platform with proof-of-work based on analog Hamiltonian optimisers}

\author{Kirill P. Kalinin$^1$ and Natalia G. Berloff$^{2,1}$ }
\affiliation{$^1$Department of Applied Mathematics and Theoretical Physics, University of Cambridge, Cambridge CB3 0WA, United Kingdom}
\affiliation{$^2$Skolkovo Institute of Science and Technology Novaya St., 100, Skolkovo 143025, Russian Federation}

\date{\today}

\begin{abstract}{The development of quantum information platforms such as quantum computers and quantum simulators that will rival classical Turing computations are typically viewed as a threat to secure data transmissions and therefore to crypto-systems and  financial markets in general. We propose to use such platforms  as a proof-of-work protocol for blockchain technology, which underlies cryptocurrencies providing a way to document the transactions in a permanent decentralised public record and to be further securely and transparently monitored. We reconsider the basis of blockchain encryption and suggest to move from currently used proof-of-work schemes to the proof-of-work performed by analog Hamiltonian optimisers. This approach has a potential to  significantly increase decentralisation of the existing blockchains and to help achieve faster transaction times, therefore, removing the main obstacles for blockchain implementation. We discuss the proof-of-work protocols for a few most promising optimiser platforms: quantum annealing hardware based on D-wave simulators and a new class of gain-dissipative simulators.}
\end{abstract}

\maketitle

Blockchain technology with its digital currency, Bitcoin, was originally introduced about ten years ago \citep{BitcoinOriginalPaper2009} and was quickly followed by the development of many other cryptocurrencies. Bitcoin was the first decentralised electronic payment system operated by an open peer-to-peer network where a financial transaction happens directly between two willing parties without the need for a trusted intermediary such as banks or other financial institutions. The deployment and enhancement of blockchain technology may completely transform the existing financial system in the next few years. The world leading financial institutions started investing in startups based on blockchain technologies, exploring its novel applications and opportunities. The major banks are establishing a framework for using the blockchain technology in the financial markets \citep{BanksUnited}. The acceptance of bitcoin in Japan together with a recently established Crypto Valley in Switzerland are the first signs that cryptocurrencies become recognised on the governing level. 

In principle, all industries that serve as intermediaries for processing financial transactions will have to adapt and compete with the blockchain alternatives. The blockchain will capture various aspects of our lives including non-financial applications such as the implementation of a decentralised platform for the Internet of Things, health records and notary, loyalty payments in the media industry, private securities. IBM and Samsung are now developing a new platform called ADEPT (Autonomous Decentralised Peer To Peer Telemetry) based on the blockchain that will keep a trusted record of all the messages exchanged between smart devices in a distributed network. Nasdaq is implementing private equity exchange on top of Blockchain with a goal to create a more secure, efficient system to trade stocks \citep{ChainCom}. DocuSign, a company that specializes in electronic contracts, just unveiled a joint idea with Visa to use blockchain to track car rentals and reduce paperwork. Microsoft develops Azure Blockchain to allow the  developers and customers to create private, semi-private, public and consortium blockchain networks with a single click on the Azure's cloud platform, thereby enabling them to distribute blockchain products. Moreover, the majority of all possible applications can be realised on the same blockchain by using the "smart contracts", which are computer programs that can automatically execute the terms of a contract \citep{SmartContract1996}. When a preconfigured condition in a smart contract among participating entities is met,  the parties involved in the contractual agreement can  automatically make the contractual payments in a transparent manner. Companies like ethereum and Codius are already enabling Smart contracts using blockchain technology and many companies which operate using blockchain technologies are beginning to support Smart contracts. Many cases where assets are transferred only after meeting certain conditions, which require lawyers to create a contract and banks to provide escrow services, can be replaced by Smart contracts. Ethereum is already powering a wide range of early applications by using Smart contracts in areas such as governance, autonomous banks, keyless access, crowdfunding and financial derivatives trading.

\begin{figure}[t!]
\centering
  \includegraphics[width=8.6cm]{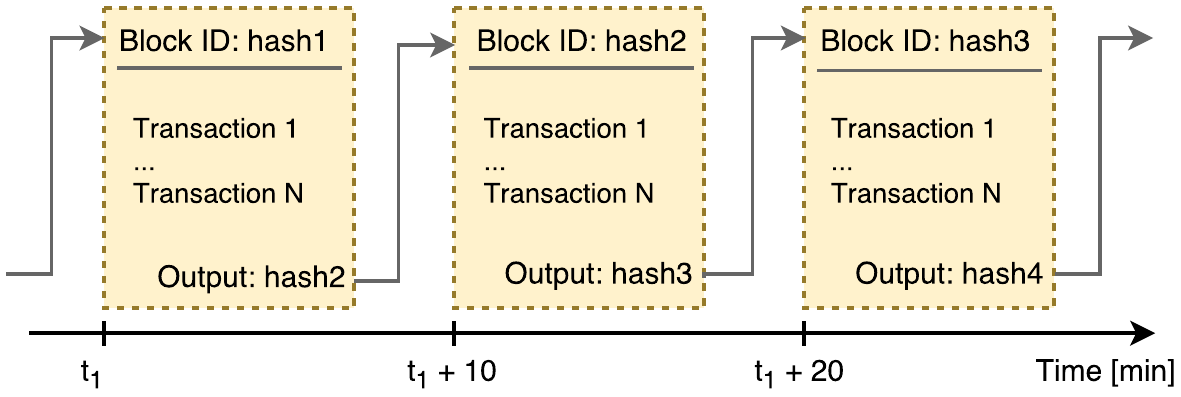}
\caption{Blockchain generation by computational nodes in Bitcoin cryptocurrency. The block is added to a blockchain with an average interval of ten minutes and it contains approximately 4000-5000 transactions. The output of each block serves as the name for the next block, thus forming a chain.}
 \label{Figure0}
\end{figure}

The  investments attracted by main cryptocurrencies caused an increased interest in understanding the structure and technological capabilities of the platform. The blockchain consists of a publicly accessible database of all transactions that are arranged into blocks of a certain length in the order preserved by a distributed ledger (so shared across multiple sites), see Fig. \ref{Figure0}. The decentralisation of the blockchain is currently insured by the distributed computational powers (computational nodes) verifying all the transactions and agreeing about what blocks should be on the blockchain, so there is no specific computer responsible for a particular transaction. In order to validate the transaction of a particular block, one can wait until several newer blocks are added to the blockchain which will automatically validate all of the previously created blocks.

A major technical obstacle, that prevents Visa and other payment systems to be replaced by digital cryptocurrencies for daily transactions, is the transaction confirmation time which is dependent on the processing time of each new block in a blockchain. In bitcoin, the functions of an intermediary, i.e. the "trust", is based on the amount of work performed. On one hand, it has to be hard enough for an  ordinary CPU to process it instantaneously (therefore the complexity of the work is controlled and at the moment is limited  to ten minutes on average to process  each block added to a blockchain). On the other hand, for routine everyday transactions the time spent per block has to be much shorter than that. This conflicting requirement prevents all of the existing cryptocurrencies from becoming a real electronic payment system that can be used for  routine financial  transactions.  Another problem is the violation of decentralisation. The $70\%$ of the cryptocurrencies with the highest capitalisation are already controlled by a few major computational nodes. Such centralised hubs of power make it impossible for a new computational node to join the system unless it matches the huge computational power of centralized nodes. Possible solutions of these problems  are seemingly mutually exclusive:  shortening of the processing time has to be accompanied by restrictions on such processing for modest computational powers, but not to lead to computational centralization!  However, we argue in this paper that a conceptually different  scheme of blockchain processing  which relies on using recently emerged analog Hamiltonian optimizers (AHO) and quantum simulators may be capable of fulfilling these requirements and revolutionarizing the blockchain technology. We also argue that the modern capabilities of such optimizers are already sufficient to implement a proof-of-work to record a public history of transactions by solving NP-hard problems that are computationally impractical for conventional classical computers and CPUs, but can be easily solved on the purpose built analog Hamiltonian and quantum simulators. Moreover, our approach does not concentrate on a particular cryptocurrency such as bitcoin and we do not limit ourselves to blockchain technology, but we rather focus on a hard computational problem that can be used as proof-of-work in future versions of any cryptocurrencies or other technologies based on proof-of-work principle. 

Our paper is organised as follows. In Section 1, we give an overview of blockchain technology and the way it drives cryptocurrency payment systems. In Section 2, we discuss two classes of hard optimization (NP-hard) problems, namely quadratic unconstrained binary optimisation  (QUBO) and constant modulus constrained quadratic continuous optimization (QCO). These problems are the type of problems that can be  solved by  AHO/quantum simulators faster than by classical computer. We discuss the D-Wave simulators based on superconducting cubits and recently emerged gain-dissipative simulators, including polariton, trapped photon condensates, and atomic multi-mode QED (quantum electro-dynamics). In Section 3, we propose how QUBO and QCO solved by analog simulators can be used in a blockchain technology as a new generation of proof-of-work protocol, which ensures much better decentralisation compared to the existing protocols and provides a greater transaction rate. Finally, we conclude in Section 4.

\section{Blockchain and Proof-of-work principle }

Adding each new block of transactions to the blockchain requires solving a computationally demanding problem, i.e. proof-of-work (POW). POW concept was originally developed to prevent junk mails by requiring the sender to solve a moderately hard computational problem to allow for the message to be delivered \citep{PoWconcept1993}. In the blockchain, POW concept is implemented by computational nodes that perform complex mathematical calculations and are rewarded for this by cryptocoins. This is often the only way for cryptocoins to enter the system, and hence the computational nodes are called miners and the process of performing the computations is called mining.

The usual POW problems are based on a function $H$, called hash function, which maps an arbitrary sized input data to a fixed size output (called hash) and is  designed to be hard to invert, i.e. the hash $y$ can be easily computed from the initial data $x$ by calculating $y=H(x)$ but finding $x$ from a given $y$ is  computationally hard. The inversion of the hash function requires an exponentially growing  computational time of an order of $O(2^n)$ where $n$ is the hash size, but when $x$ is found the validation of the transaction could be easily done  by computing  $H(x)$ and comparing the result with the hash $y$. Every transaction in the block has a hash associated with it and each block in the blockchain is identified by its block header hash (see Supplementary Material for the full list of parameters included in it).

The mining difficulty for adding new block, which is represented by  the difficulty  target value, is dynamically controlled and regularly adjusted by a moving average giving  an average number of blocks per hour fixed  in order to compensate the increasing computational power and varying interest in running nodes involved in mining. In bitcoin, the difficulty target value is updated every 2016 blocks in order to target the desired block interval accurately, which is now set to be ten minutes. This rate  is chosen as an ad hoc tradeoff: the time which is too  short  would decrease the stability of the blockchain (more forks and longer forks in the blockchain tree would require an increased bandwidth between nodes); the time which is too long  would increase the confirmation time of transactions. Other cryptocurrencies have different times per block, e.g. Litecoin has a 2.5 minutes block time. 

\textit{Centralisation problems.} The initially supposed global decentralisation of the bitcoin and other cryptocurrencies is now under risk since the computational nodes tend to unite in computational pools followed by emergence  of major cryptocurrencies with highest capitalisation on the market, i.e. bitcoin and etherium, which are controlled by 5-7 different parties. The network remains decentralised but with a few centralised hubs. This in turn leads to a high entry barrier for a new potential mining party since it has to compete with the computational rate of centralised hubs, which will make the system even more centralised in the nearest future.
For a brief review of security issues and other problems including energy consumption, storage of the information, please see the Supplementary Material.

In the next Section we discuss the type of hard problems that can be solved -- mined -- faster on  specially built alternatives to classical computers, namely, analog Hamiltonian optimisers/quantum simulators. We propose to put such problems and such solvers at the core of POW schemes that the agents need to solve in order to add each new block of transactions to the blockchain.

\section{Hard computational problems and their solvers for blockchain POW}
At the heart of Richard Feynman's idea of quantum simulator lied the proposal of using one well-tunable quantum system to simulate another quantum system \cite{feymann82}. To design  such quantum simulator \citep{BlochNature2012}, one needs to map the variables of the desired Hamiltonian of the system into the elements (spins, currents etc.) of the simulator, tune the interactions between them, prepare the simulator in a state that is relevant to the physical problem of interest, and perform measurements on the simulator  with the required precision.  In the past decade, this original meaning of quantum simulator has been widened and modified to include the platforms that intend to solve  classical optimization problems faster than classical computer for a given problem size (in terms of the number of variables, and therefore, the dimensionality of the function to be optimized). Various physical systems have been proposed and realised as such simulators to a various extent \cite{georgescu14}. Among those are systems that use essentially quantum processes for their operation  (e.g. entanglement and superpositions)  such as  trapped ions \cite{kim10,lanyon} or  superconducting qubits \cite{corcoles}, for others although the quantum processes are crucial in forming the state of the system such as  Bose-Einstein condensates, the system behaves as a classical system, e.g  ultracold atoms in optical lattices \cite{reviewUltracold,saffman,simon11,fermionic}, network of optical parametric oscillators (OPOs) \cite{yamamoto11, yamamoto14},  coupled lasers \cite{coupledlaser}, polariton condensates \cite{NatashaNatMat2017}, multimode cavity QED \cite{qed} and photon condensates \cite{photon}. These systems emulate spin Hamiltonians such as Ising, XY or Heisenberg (so called $n-$ vector models).  Hardness of the problem depends on the number of nodes (eg. bits, qubits), on the ability to control couplings between the elements and the overall connectivity of the system. The  existence of  universal spin Hamiltonians has been established.  Universality means that all  classical $n$-vector  models with any range of interactions can be reproduced within such a model, and certain simple Hamiltonians such as  2D Ising model on a square lattice with transverse fields are  universal \cite{CubittScience16}. Indeed, such problems are NP-hard for a general matrix of couplings -- the number of operations grows as an exponential function with the matrix size. This suggests that one can formulate a spin Hamiltonian (Ising, XY or Heisenberg) for which the global minimum can be found by a simulator optimized for solving such problems  much faster than on classical computer. 

Finding the optimal solution of the general $n$ vector model for a sufficiently large size  may be suitable for a POW protocol. Here we discuss two of such problems as the system requirements for simulators mentioned above designed to solve these problems.  First of these is the quadratic unconstrained binary optimisation (QUBO) problem which is an optimization formulation of a max-3-cut problem for vector ${\bf z} \in \mathbb C^N$ with components $z_i, i=1,\cdot\cdot\cdot N$ and $N\times N$ real symmetric matrix $Q$
\begin{equation}
\max {\bf z}^H Q {\bf z}, \quad {\rm subject}\quad {\rm to} \quad z_i\in \{-1,1\},
\label{qubo}
\end{equation}
and, second, quadratic continuous optimisation (QCO) problem 
\begin{equation}
\max {\bf z}^H Q {\bf z}, \quad {\rm subject}\quad {\rm to} \quad |z_i|=1.
\label{qco}
\end{equation}
 QUBO is a discrete version of QCO for which the decision variables are constrained to lie on the unit circle,which is a continuous domain. The Ising model
\begin{equation}
\min \quad -\sum_{i<j} J_{ij} s_i s_j \quad {\rm subject}\quad {\rm to} \quad s_i \in \{-1,1\}
\label{ising}
\end{equation}
and XY model
\begin{equation}
\min \quad -\sum_{i<j} J_{ij} {\bf s}_i \cdot {\bf s}_j \quad {\rm subject}\quad {\rm to} \quad s_i =(\cos \theta_i, \sin\theta_i), 
\label{xy}
\end{equation}
 are trivially mapped into QUBO and QCO, respectively, by associating the "spins" $s_i$ and ${\bf s}_i$ with $z$ (via $z_i=\cos \theta_i + i \sin \theta_i$ for the XY model, and $z_i\in \{-1,1\}$ for the Ising model) and the coupling strengths $J_{ij}$ between the spins with the matrix elements of $Q$.
 
These problems are known to be strongly NP-hard in general  and, therefore, even  for medium-sized instances are difficult to solve on a classical computer \cite{ZHANG06}. The time required to find the solution depends on the matrix structure: its size, number of zero entries (sparsity), the way the elements are generated, whether it is positive-definite or indefinite matrix etc.   The algorithms for solving these problems on a classical computer can be divided into three types:  exact methods that find the optimal solution to the machine precision, approximate algorithms that guarantee that the solution will be found within some approximation ratio and heuristic algorithms where suitability for solving a particular problem comes from empirical testing \cite{empirical}. Exact methods  typically involve a tree search of a general branch-and-bound nature and the exponential worst-case runtime. They can be used to solve a limited range of problems for small or sparse matrices.  The heuristic algorithms such as  simulated annealing, genetic and evolution algorithms can deliver a good, but suboptimal (and possibly infeasible) point in a  short amount of time \cite{wang14}.   Approximate algorithms find an approximate solution -- an optimal value of $z^HQz$, which is at least a constant times the true optimal value. Such constant is called a "performance guarantee" of the algorithm. Both QUBO and QCO are known to be  APX-hard problems \cite{APX}  meaning that there is no polynomial-time approximation algorithm that gives the value of the objective function that is arbitrarily close to the optimal solution (unless P = NP). For these problems, therefore, the perfomance guarantee is bounded by a constant and no approximate algorithms can be devised  to do better. The approximate algorithms are typically based on some form of semidefinite programming relaxation (SDP). The achieved performance guarantee depends on the structure of the matrix. For instance, for positive semidefinite $Q$ with the elements of the same sign the  performance guarantee of SDP methods can be as high as $\frac{2}{\pi} \min_{0\le \tau\le \pi} \frac{\tau}{1-\cos \tau}\approx 0.878$ \citep{GW04}, however, if the assumption about the sign of elements  is relaxed the performance guarantee becomes 0.537 for QUBO and $\pi/4$ for QCO \citep{ZHANG06} . An approximation algorithm on a more general indefinite matrix will yield even worse approximate. Furthermore, the task of  finding an  approximation for maximizers ${\bf z}$  instead of the approximation to the objective function ${\bf z}^H Q {\bf z}$  easily becomes non-computable \citep{hansen}. 
The best classical computational algorithms capable of finding the solution of the QUBO/QCO problems for a general matrix $Q$ are limited to very modest sizes. For instance, for $N=200$ with only 30\% nonzero elements the state-of-the art algorithms would take on average 80 minutes to solve QUBO \citep{bcrunch17}.

Since the spin Hamiltonian models are straightforwardly mapped into QUBO and QCO, it is natural that analog/quantum simulators based on condensed matter quantum systems have architecture most suitable for solving such problems. Below we consider the most promising platforms for solving QUBO or QCO.

\textbf{Quantum Annealers.}  D-Wave is a first commercially available quantum annealer that is built on superconducting qubits with programmable couplings and specifically designed to solve QUBO problems (Eq. \ref{qubo}) \citep{DwaveNature2011}. By specifying the  interactions $J_{ij}$ between qubits, a desired QUBO problem is solved by quantum annealing process \citep{QuantAnneal}. Adiabatic  (slow) transition in time from an initial state of a specially prepared "easy" Hamiltonian to the objective Ising Hamiltonian  guarantees that the system remains in the   ground state, which gives the final energy that corresponds to the optimal solution of the QUBO problem. 

First benchmarks on random QUBO problems were performed on a "D-Wave One" and "D-Wave Two" questioning the fact of quantum speedup of annealer \citep{QuanAnnealPerfomanceTroyer2015}.  It was demonstrated that although the D-wave One simulator (with 128 qubits) is a true  quantum annealer it is not yet competitive with classical computing technology. No speedup was found for  problems of  sizes ranging from 8 to 512 qubits. Later, it was shown  \citep{DwavePRX2015}, that for carefully crafted problems with rugged energy landscapes that are dominated by large and tall barriers, the quantum annealer can offer a significant runtime advantage over a classical version of simulated annealing. For some problems with sizes involving nearly 1000 binary variables, quantum annealing was up to $10^8$ times faster than classical simulated annealing (SA) or Quantum Monte-Carlo (QMC)  methods running on a single core CPU. The quantum speedup was not claimed again since a variety of heuristic classical algorithms can solve most instances of  problems structured on Chimera architecture of  D-Wave computers more efficiently \citep{HeuristicAlgo}.

These results do not exclude a possible quantum speedup for some other problems that can be specifically created to suit the machine's abilities and  which will benefit from quantum anealing. With the last commercially available  D-Wave machine released in 2017, quantum annealers may finally outperform all classical algorithms. For evaluation of the new 2000-qubit D-Wave QPU a new synthetic problem class was proposed \citep{DwaveTeamArXiv2017} with more emphasis on creating computational hardness through frustrated global interactions. Such frustration creates meaningful difficulty for general heuristic algorithms that are unaware of the planted problem class. The D-Wave team claimed the 2000Q could find solutions up to 2600 times faster than any known classical algorithm \citep{DwaveTeamArXiv2017}. This time the D-Wave simulator was competing with the state-of-the-art CPU implementations of SA, QMC, support vector machine classification algorithm, and Selby's CPU implementation of Hamze-de Freitas-Selby algorithm, making the competition much stronger than it was in \citep{DwavePRX2015}. Three orders of magnitude speedup over software solvers, reported for pure annealing time (computation time), translates into a 30 times speedup in total wall clock time including programming and readout necessary for D-Wave machine. 

One of the major limitations of D-wave simulators is that each qubit can be connected to maximum of six other qubits and thus $N^2$ qubits are needed for encoding $N$-variable problem. The next generation of D-Wave quantum computer is expected to be announced in 2018 with new powerful capabilities such as reverse annealing and virtual graphs. These features are expected to enable significant performance improvements over the current D-Wave simulators by giving users increased control of the QPU and will probably make the fact of a quantum speedup even more clear.

\textbf{OPO-based simulators.} Network of coupled optical parametric oscillators (OPOs) is an alternative physical system for solving the Ising problem (\citep{YamomotoScience2016} and references therein). A scalable optical processor with electronic feedback  was realized for solving Ising problem with up to 100 spins and 10,000 spin-spin connections \citep{YamomotoScience2016}. In this Coherent Ising Machine (CIM), the ground state of the Ising Hamiltonian corresponds to an oscillation mode with the minimum network of degenerate OPOs loss. The fully programmable connections between any two spins is a significant difference of this model compared to the D-Wave simulator.  The first promising signs of speedup of CIM have been reported recently \citep{YamomotoNPJ2017}, however, the fundamental computational power of OPO Ising machines or the time required to program  thousands of connections and to readout of the final state have not been fully explored.

\textbf{Gain-Dissipative Simulators.} We define gain-dissipative simulators as the optimisers based on a driven-dissipative physical system. The principle of their operation depends on the gain (pumping)  mechanism.  As the gain exceeds some threshold a coherent state of matter emerges maximising the state occupation and therefore solving QUBO or QCO \citep{NatashaNatMat2017}. These are novel systems and their potential as analog simulators has been very recently demonstrated \citep{DavidsonPRL2013,NatashaNatMat2017, qed,photon}. These platforms could be referred to as quantum simulators due to the quantum-statistical nature of the formation of the coherent state (e.g. Bose-Einstein condensate) during which a large fraction of bosons occupies the lowest quantum state and the macroscopic quantum phenomena emerges.  These systems enjoy a quantum speed-up which is associated with the stimulated process of condensation i.e. an accelerated relaxation to the global ground quantum state. However, after the condensate formation the system behaves classically as non-commutativity of field operators becomes insignificant in the large number of particles regime. To distinguish these platforms from quantum computers/quantum simulators that rely on entanglement and quantum superposition, we refer to them as "analog Hamiltonian simulator/optimizer" (AHO). The search for the solution of QUBO or QCO is via a bottom-up approach which has an advantage over classical or quantum annealing techniques, where the global ground state is reached through either a transition over metastable excited states or via tunnelling between the states in time that depends on the size of the system. Different AHOs rely on different quasi-particles as the basis for bit/qubit and vary by scalability, coupling control, connectivity and the accuracy with which the result can be read, etc. The microscopics of various systems we present below can be quite different, but the governing principle for solving QCO is based on the representation of each "spin"  indexed by $i$ and  centered at the position ${\bf x}_i$ by the wavefunction $\Psi (|{\bf x}-{\bf x}_i|) \exp[{\rm i} \theta_i]$, where $\Psi$ is the wavefunction of an isolated condensate centered at the origin. At the condensation threshold the phase differences between the individual wavefunctions  maximize the total number of particles given by
\begin{eqnarray}
{\cal M}& \equiv &\int \big |\sum_i \Psi (|x - x_i|) \exp[{\rm i} \theta_i]\big|^2\, d{\bf x} \nonumber \\
& =& N \int \big|\Psi \big|^2 \, d{\bf x} + \sum_{j<i} J_{ij} \cos(\theta_i-\theta_j),
\label{M}
\end{eqnarray}
where $J_{ij}=\int [\Psi(|{\bf x}-{\bf x}_i|)\Psi^*(|{\bf x}-{\bf x}_j|) +c.c]\, d{\bf x}$. Since the first term on the right hand side does not depend on the phases, the maximization of ${\cal M}$ is equivalent to minimization of Eq. (\ref{xy}), and, therefore, to finding the solution to QCO. We conclude that such gain-dissipative platforms solve QCO at the condensation threshold \citep{NatashaNatMat2017}. To understand the formation dynamics of such driven-dissipative system one can replace the spatially dependent condensate profile centered at ${\bf x}_i$ by a complex function  $\psi_i(t)$ with the dynamics described by a rate equation \cite{PO}
\begin{equation}
\dot{\psi_i} = (\gamma_{\rm eff}^i -i v^i- \sigma |\psi_i|^2-i U|\psi_i|^2)\psi_i + \sum_{j} J_{ij} \psi_j,
\label{rate}
\end{equation}
where  $\gamma_{\rm eff}^i$ is the effective gain given by the difference between the slowly increasing function of pumping and a constant linear dissipation at site $i$, $v^i$ represents the blue shift due to the external potential applied at the site $i$,  $\sigma$ corresponds to the nonlinear losses that allow for gain saturation and therefore, for a steady state at the condensation threshold, and $U$ describes the strength of nonlinear interactions at the site $i$. If one writes $\psi_i$ in terms of the density $\rho_i(t)$ and phase $\theta_i(t)$ using $\psi_i=\sqrt{\rho_i} \exp[{\rm i} \theta_i(t)]$ and separates real and imaginary parts of Eq. (\ref{rate}) then the resulting equations yield
\begin{eqnarray}
\frac{1}{2}\dot{\rho_i} &=& (\gamma_{\rm eff}^i - \sigma \rho_i)\rho_i +  \sum_{j} J_{ij} \sqrt{\rho_i \rho_j} \cos\theta_{ij} +\chi_i \label{1eq}\\
\dot{\theta_i} &=& -v^i-U\rho_i - \sum_j J_{ij} \frac{\sqrt{\rho_j}}{\sqrt{\rho_i}}\sin\theta_{ij} + \xi_i,
\label{2eq}
\end{eqnarray}
where $\theta_{ij}=\theta_i-\theta_j$ and we included a small density and phase perturbations $\chi_i(t)$ and $\xi_i(t)$ that respectively represent the spontaneous and stimulated scattering during the condensation process, incorporate classical and quantum effects and disappear at the condensation threshold. It follows from Eqs. (\ref{1eq}-\ref{2eq}) that the gradient flow to the solution of QCO is realised if all $\rho_i$ are the same, which is achieved by adjusting $\gamma_{\rm eff}^i$ during the condensate formation.  At the condensate threshold the steady state is achieved with $\dot{\rho_i}=0$ and  $\dot{\theta_i}=\mu={\rm const}$, where $\mu$ is the chemical potential of the system. To drive the system to such a state the adjustment of the pumping of the individual nodes has to be accompanied by the adjustment of the external potentials $v_i$ \cite{kirillfuture}.

\begin{figure}[t!]
\centering
  \includegraphics[width=8.6cm]{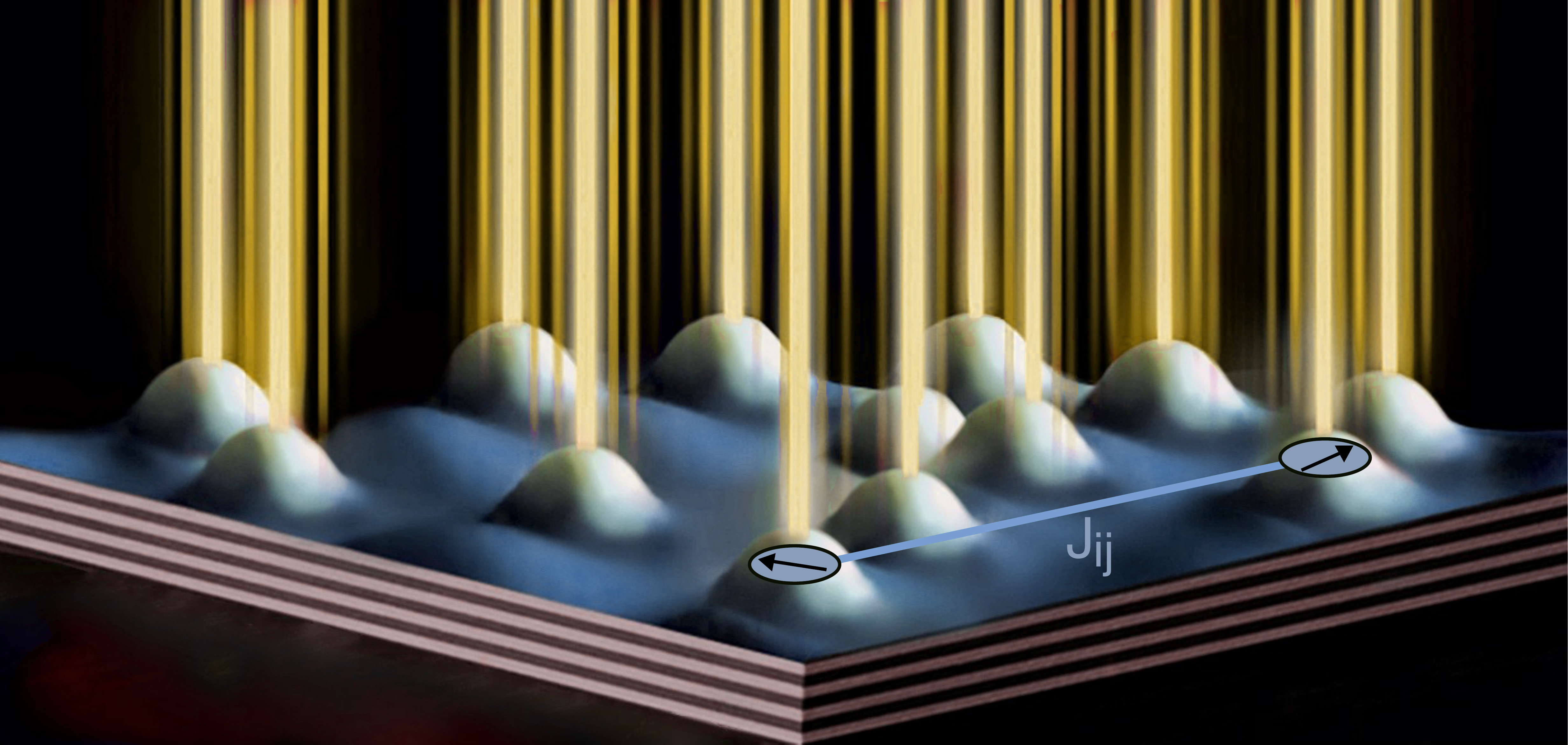}
\caption{A general scheme for a gain-dissipative simulator. The condensates are pumped in a two-dimensional lattice with interaction strengths $J_{ij}$ between the adjacent vertices. The final phase configuration minimizes the $XY$ or $Ising$ Hamiltonians depeding on a particular physical system. }
 \label{Figure1}
\end{figure}

Next we discuss the the actual physical platforms representing the gain-driven analog Hamiltonian optimizers.

\textbf{Polariton graph optimizers.}
Polariton graph optimizers are  based on exciton-polariton condensates arranged at a particular graph geometry  and are used for solving QCO (Eq. \ref{qco}).  Exciton-polaritons (or polaritons) are the composed light-matter  quasi-particles formed in the strong exciton-photon coupling regime in semiconductor microcavities \citep{ExcPol1992}. At low densities these quasi-particles are bosons and condense above some threshold of pumping intensity forming a single coherent state. Polariton condensates can be imprinted into any two-dimensional graph by spatial modulation of the pumping laser. Such two-dimensional graphs  of condensates inherit high flexibility in engineering any geometrical configuration of nodes. By controlling the excitation density, the profile of the pump, the graph geometry and the separation distance between the lattice sites, one can control the couplings between the sites  and realise various phase configurations between individual condensates. At the condensation  threshold the phase differences, $\theta_i-\theta_j$, between condensates at different nodes indexed by $i$ and $j$ establish the minimum of the XY model (\ref{xy}) and, therefore,  solve  QCO \citep{NatashaNatMat2017,PO}. Polariton graphs are easily scalable and the graphs that consist of 45 and 100 nodes have already been realised \citep{NatashaNatMat2017}.  The coupling strengths $J_{ij}$ 
in (Eq. \ref{xy}) and so the elements of $Q$ in (Eq. \ref{qco}) are controlled by the graph geometry and go beyond the next neighbour interactions \citep{Exotic17}.

\textbf{Photon-based simulators.}
Photon condensates as the elements (bit/qubit) of AHO for solving QCO can be  created by generating variable potentials for light within an optical high-finesse microcavity \citep{KlaersNatPhotonics2017}.  The long photon lifetime enables the thermalization of photons and the demonstration of a microscopic photon condensate in a single localized site. The investigation of effective photon-photon interactions as well as the observed tunnel coupling between sites makes the system a promising candidate for AHO. The scalability of hundreds of condensates has been already demonstrated \citep{KlaersNatPhotonics2017} suggesting that thermo-optic imprinting provides a new approach for variable microstructuring in photonics.

\textbf{QED-based simulators.} 
A realization of a multimode cavity QED (quantum electrodynamics) system  \citep{BenLevPRX2018} paves a way to a QED-based AHO thanks to the strong, tunable-range, and local interactions between Bose-Einstein condensates trapped within the cavity. While single-mode cavities provide strong but infinite-range photon-mediated interactions among intracavity atoms, it was experimentally shown in \citep{BenLevPRX2018} that local couplings can be created using multimode cavity QED. Moreover, atom-atom couplings may be tuned from short range to long range which reduces the sparseness of matrix $Q$ and therefore increases the potential complexity of the problem. 

The XY model has been previously simulated by other physical systems: ultra cold atomic optical lattices \citep{StruckScience2011} and coupled photon lasers network \citep{DavidsonPRL2013}, which was also proposed for solving  QCO. 
 We have not discuss here other universal fully quantum platforms such as based on superconducting qubits \citep{MartinisArxiv2017} or trapped ions \cite{trappedIons} as their potential for solving QUBO and QCO is not fully explored.

\section{Laying out analog Hamiltonian simulators on a Blockchain.}
In the previous section we considered several possible analog Hamiltonian simulators that start being compared to classical state-of-the-art algorithms for solving the global optimization problems such as QUBO and QCO. These platforms, have either demonstrated a speedup, or have a potential to achieve this   in the nearest future. Figure \ref{Figure1} illustrates the schematics of the POW protocols that can be based on solving QUBO or QCO problems using the currently available analog Hamiltonian simulators. 

\begin{figure}[b!]
\centering
  \includegraphics[width=8.6cm]{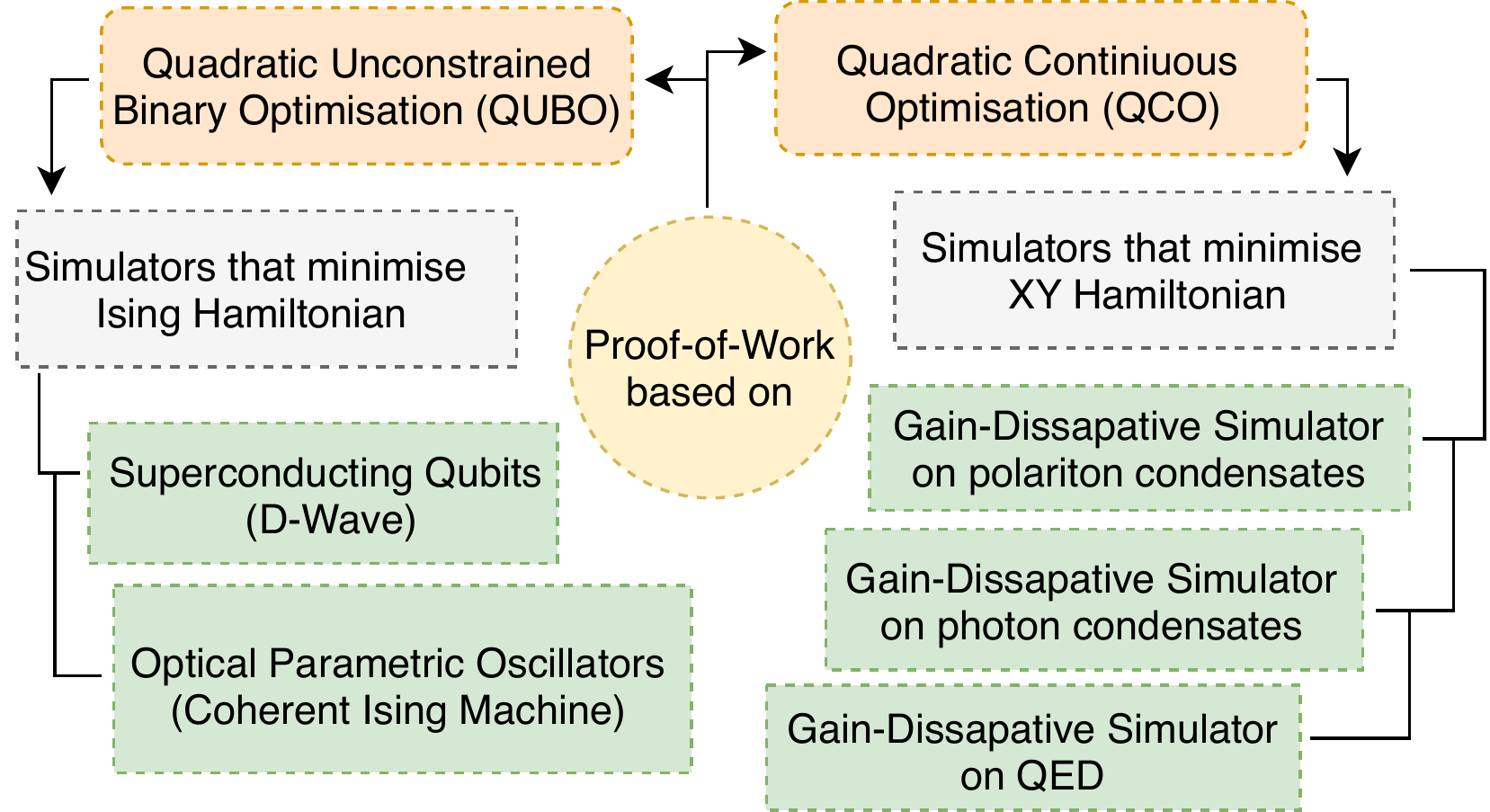}
\caption{The scheme shows that the Proof-of-work protocol can be realised for solving the QUBO or QCO problems on purposely built quantum simulators based on superconducting qubits, OPO-s, polaritons, photons, QED. }
 \label{Figure1}
\end{figure}

The recipe for building a blockchain based on such simulators is essentially the same for all types of simulators. The input for each block will include among  other parameters (like timestamp, previous block id, etc.) a parameter that is specific for each type of simulator. In case of the simulators considered here, this additional parameter is a matrix of coupling strengths $J_{ij}$ (the elements of matrix $Q$). The ways of controlling and modifying the coupling strengths  are system dependent,  for instance, by changing the distance, pumping intensity and the hight of the trap barrier in  gain-dissipative simulators. The coupling matrix has to be formed depending on the block content, so nobody can prepare a coupling matrix in advance and solve it in order to approve a particular block. For instance, the numerical expression of the order of transactions together with the amount of each transaction could be used to form this matrix. The output of the block consists of the problem optimizers: the resulting "spins" $s_i$ or phases $\theta_i$, that can be further encoded and serve as a next block's name. Verifying whether the block belongs to the chain or not can be done  for instance by checking that the value of the objective function ${\bf z}^H Q {\bf z}$ for the found optimizers ${\bf z}$ is larger (better) than the one found by a classical approximation or heuristic algorithm.

 Next we explicitely construct the matrix $Q$. Suppose that a blockchain consists of blocks with  capacity of $X$  transactions per block. Depending on the capabilities of a particular simulator/optimizer a different range of coupling strengths $J_{ij}$ can be realised (for instance, a polariton graph optimizer can achieve the coupling strengths between $-2 \mu eV$ to $2 \mu eV$ \cite{Mateo17}). Based on the platform we  define a hash function $H_0(x)=y$  that maps a transaction $x$ into an output  $y$ where $y\in [\min {J_{ij}},\max {J_{ij}}].$  To generate the coupling matrix $Q$ of size $N\times N$ and sparsity  $(100-D)\%$  we need to map the block content  into $m=N(N-1)D/200$ non-zero matrix elements. If $X=m$, each transaction is hashed into a non-zero matrix element directly using $H_0$. If  $X>m$  a necessary number of transactions has to be hashed together (by a different hash $H$ that returns the same size of the output as input) before mapping using $H_0$ is performed. If $X<m$ then the required  number of non-zero elements  is generated by first hashing all transaction individually, then adding the hashes of their pairs, and so on until the required number of elements $m$ is reached. 

For instance,  for bitcoin, the amount of transactions per block is $X \approx 5000$. The most recent D-Wave simulator has a coupling matrix of size $N \approx 2000$ which is quite sparse with $D \approx 1 \%$ which gives about 20000 non-zero matrix elements. We fill the first 5000 elements by hashing each of 5000 individual transactions, than  by hashing the hashes of the pairs of the  first, second and  third transactions with all the rest. 

Finally, we note that not only the transaction should be mapped to the coupling matrix, but the other header parameters of the block as well (timestamp, the previous block's id, see Suppl. Mat. for a full list of parameters).

\section{Conclusions}

In this article, we propose to use  the analog Hamiltonian optimisers as a basis for a proof-of-work protocols. Such simulators are capable of outperforming classical computers on the timescale of seconds or less as compared to hours or more, which offers straightforward benefits. In a particular case of blockchain technology, such optimisers will allow a faster verification process reducing the time interval between blocks  which will significantly decrease the confirmation time of transaction to the order comparable to Visa and even beyond, and a possible new generation of digital cash will be applicable on day-to-day basis. A specific achievable timescale depends on the nature of simulator and the type of the problem. In case of a commercially available D-Wave machine, for instance, proof-of-work will take tens of milliseconds. Consequently, the analog simulators considered in this article will make possible for a blockchain to be operated at a maximum synchronisation speed between the nodes. 

One can think about quantum simulators/analog Hamiltonian optimizers as the computational black boxes with approximately the same "hashing power" which is higher than the power of any classical computer. Distributing such blackboxes between 20-50 independent nodes would make the system much more decentralised than any other existing platforms of current blockchain technology. The proof-of-work protocol based on analog Hamiltonian simulators would allow better scalability by making faster the  processing of databases (sharding implementation) while current proof-of-work schemes do not support it due to security issues.

\section*{Supplementary material}

\textbf{The block header.}
The block header consists of the bitcoin version number, the hash of the previous block header, the  hash of all the hashes of all the transactions in the block  (the Merkle Root), the timestamp, the difficulty {\it target}  (the precision of calculation needed to meet the required level of POW in under ten  minutes) and a random 32-bit integer number called 'nonce' (the value that is altered by the miners to try different permutations to achieve the difficulty level required).

 \textbf{Bitcoin security.} 
The security of bitcoin is based on the two cryptographic protocols that prevent it from being stolen or copied.

The first is the POW that is required to write transactions to the bitcoin digital ledger. If a majority of computations are  done independently, the honest chain will grow faster and outpace any competing chains since all the new blocks have to be added to the longest chain. To modify a past block, an attacker would have to redo the POW of the block and all blocks after it and then catch up with and surpass the work of the honest nodes. The probability of a slower attacker catching up diminishes exponentially as subsequent blocks are added. So it's not possible to change the information about any transactions in the blocks that has been already incorporated in the blockchain. Moreover, nobody can prepare a malicious block in advance, since the result of the hash function will depend on the previous block's hash. This is why a malicious node has to compete in a computational power with the whole network of peers  in real time with the only chance to win by exceeding the total power of all other nodes. Thus, the higher total computation power of the distributed network of nodes ensures the safety of the blockchain from creating an alternative history of transactions by a particular malicious node.

Bitcoin has another cryptographic security feature to ensure that only the owner of cryptocoins can spend them which is based on an elliptic curve digital signature algorithm (ECDSA). It is based on assumption that finding the discrete logarithm in a cyclic subgroup of a random elliptic curve over a finite field is infeasible  \citep{ECDSA2001}. Following this signing algorithm, the coin owner generates two random numbers for each transaction: a private key which only the owner knows  and a public key which is revealed to the network. Only the owner is able to produce valid signatures based on these keys since the public key can be easily generated from the private key but not the other way around. In this scenario, the owner signs the hash of the transaction with the private key, and everyone else is able to validate the signature with the available public key by looking at the given signature and transaction. In this way, a signature can be used to verify that the owner possesses the private key and therefore has the right to spend the bitcoin, without revealing the actual private key.

\textbf{Security Issues.}
Knowing the two main security features of bitcoin leaves one with the two possibilities to prevail over the bitcoin's system: either controlling more than 50\% of the computational power on the network or cracking the cryptographic signature scheme.  For the latter case, the only way to cheat ECDSA is to calculate the private key using the public key, which is extremely hard problem with classical computers though its exact complexity class is not known. Quantum computers running the Shor's algorithm for number factorization \citep{Shor1994}  pose a risk to the  encryption schemes such as RSA,  and therefore to cryptocurrencies in particular. However, to break the relevant RSA codes (say RSA-1024) requires $10^5$ qubit quantum computer which is beyond the current capabilities of about 50 qubits \citep{IBM50qubitsNature2017}.

At present, there are suggestions of countermeasures that can be taken to protect bitcoin from a possible future attack of quantum computers. Examinations of how quantum computers could undermine and even exploit bitcoin security protocols have been recently discussed in \citep{BitcoinQuantAttacks2017}. Another proposal \citep{qBitcoin2017} positions transmission of quantum cryptographic keys between a remitter and a receiver of the eponomous named cryptocurrency, qBitcoin, where the exchanging qBitcoins requires a transmission network in place that can send and receive bits of quantum information, qubits. The alternative signature schemes that are believed to be quantum safe, are also discussed, and thus qBitcoin security may rely on a quantum digital signature in future.  Quantum-cryptographic improvements to the current cryptographic schemes that may be widely used in digital currencies were suggested as well \citep{QuantMoney2010,qBitcoinQMech2016,qBlockchain2017}.

\textit{Energy consumption.} In a PoW system, large amounts of electricity are required in order to power the computing hardware: the bitcoin's current annual electricity consumption is about 30 TWh which is close to consumption of such countries as Ireland, Buhrain, Slovak Republic, Oman, Morocco.   The bitcoin miners spend around $\$50,000$ per hour on electricity which is equivalent to $\approx \$450$ million per year.

\textit{Storage problem}. Up to now, bitcoin supports about $\sim$ 5-7 tps with a 1 megabyte block limit. Assuming the unlimited block size and 300 bytes per bitcoin transaction on average, it would require nearly 0.5 gigabytes per bitcoin block, every ten minutes on average to reach an equivalent capacity of Visa transaction volume of average order of $\sim 2000$ tps (56,000 tps at peak) or Paypal of the order of $\sim 200$ tps (600 tps at peak). Continuously, that would be over hundred terabytes of data per year. Clearly, achieving Visa-like capacity on the Bitcoin network is not feasible today since this amount of data has to be stored not in one place but in all the nodes constituting the decentralised computational network. To fix the issue with the confirmation time and the number of possible transactions per second, bitcoin lightning network has been proposed, however, one of the   consequences of its implementation could be  further unwanted centralisation \citep{LighningNet_centralised}.

\textit{Alternative to proof-of-work: proof-of-stake.} In POW miners invest computing power competing for a chance to add next block to a Blockchain and win a reward for doing so. To decrease the energy demands, an alternative to POW known as the proof-of-stake (POS) scheme was suggested making the entire mining process  virtual: individual computational nodes reach consensus about each next block by betting their money.  The stakeholders who will form the next block are randomly selected from a pool of validators with respect to the size of stake they have, and once the block is added to a Blockchain, a reward proportional to the stake is given. Betting on the wrong or the malicious block is to be punished by the system, which is, in case of etherium's POS suggested protocol the entire invested stake may be lost \citep{EtheriumProtocol}.

For a blockchain to be secure, the means of selecting a stakeholder to make a block must be truly random. Ouroboros introduced the randomness into the leader election process  by way of a secure, multiparty implementation of a coin-flipping protocol which is realised in Cardano cryptocurrency \citep{PoS2017}.


\end{document}